# A Java Based Architecture of P2P-Grid Middleware

Hudzia B., Ellahi T.N., McDermott L. and Kechadi T.
School of Computer Science and Informatics
University College Dublin
Dublin - Ireland
{benoit.hudzia, tariq.ellahi, liam.mcdermott, tahar.kechadi}@ucd.ie

*Abstract*—During the last decade there has been a huge interest in Grid technologies, and numerous Grid projects have been initiated with various visions of the Grid. While all these visions have the same goal of resource sharing, they differ in the functionality that a Grid supports, characterization, programming environments, etc. We present a new Grid system dedicated to dealing with data issues, called DGET (Data Grid Environment and Tools). DGET is characterized by its peer-to-peer communication system and entity-based architecture, therefore, taking advantage of the main functionality of both systems; P2P and Grid. DGET is currently under development and a prototype implementing the main components is in its first phase of testing. In this paper we limit our description to the system architectural features and to the main differences with other systems.

## I. INTRODUCTION

In recent years, Internet-scale systems have been developed and deployed to share resources at a very large scale across the traditional organisational boundaries. The need for constructing such systems was motivated by the increasingly complex requirements of modern applications from diverse disciplines. Such global scale systems provide opportunities to harness idle resources which are distributed and heterogeneous. Another benefit offered by such systems is that they allow coordinated use of resources from multiple organisations. Thus, these wide-area systems may span multiple organisations and form virtual organisations on top of the existing organisational hierarchies. Two such systems exploiting these views include Grid and Peer-to-Peer (P2P) systems. Grid and P2P systems have seen a rapid evolution and widespread deployment. The two technologies appear to have the same final objective, pooling and coordinating large sets of distributed resources[1]. During the last few years various projects have been undertaken to try to merge these two complementary technologies, such as OurGrid[2]. Also various modifications to the Globus toolkit[3] have been proposed to include P2P technology and thus improving the discovery system[4]. Typically, Grid systems are designed to run applications with intensive computing and storage needs across the traditional organisational boundaries[5], [6], [7]. They are characterised by their sophisticated resource management and data transfer components. P2P systems on the other hand were mainly designed for resource sharing, mostly files. Therefore, the focus of P2P systems is on providing sophisticated resource discovery capabilities. Both approaches have their own advantages and disadvantages.

In this paper we present the DGET (Data Grid Environment and Tools) [8][9][10] project undertaken by the Parallel Computational Research Group at UCD. DGET middleware employs techniques from both Grid and P2P systems. The rest of the paper is structured as follows: Related work is discussed in Section 2 followed by DGET concepts and architectural overview in Sections 3 & 4 respectively. Implementation is explained in Section 5. Section 6 describes future directions of the project.

## II. RELATED WORK

DGET is a P2P Grid middleware and employs techniques from both fields and as such should be compared to other middlewares adopting the P2P Grid approach.

*DGET and Grid Middleware:* A number of Grid middlewares has been developed and used. A wide range of systems have been developed. Some of these focus on providing the core middleware services while other programming frameworks are built on top of these middleware systems and provide high-level application development functionalities. Globus, Legion and UNICORE are the most notable Grid middlewares. The Globus Toolkit is the most widely used middleware. DGET has some distinct characteristics. First, existing Grid middlewares adopt a manual and static topology whereas DGET is based on a dynamic, self-organizing topology borrowed from the decentralised P2P systems. Other distinguishing DGET features include decentralized P2P style resource discovery and fine grained access control. Existing Grid systems depend on specialized central servers to maintain information about shared resources. DGET, on the other hand adopts the P2P style decentralized resource discovery approach and thus doesn't rely on any specialized servers.

On the security front, Globus possess an extremely powerful security system but it has considerable management overhead. All the users are required to have individual accounts on the resource before they can use the resource. This situation is applicable if there are a limited number of participants. In a situation where a very large number of users are present this technique would become very cumbersome. DGET on the contrary doesn't require users to have individual user accounts on the resources. DGET's security mechanism is based on an extended Java security model. Other aspects where DGET security differs from Globus are the fine-grained access control policies and the resource quota control. DGET uses XACML[11] to define fine-grained access control policies.

*DGET and P2P Systems:* The second class of system we can compare DGET with are P2P systems. DGET bears some similarities with P2P systems but uses a significantly different approach in other aspects.

Most of the P2P systems share file resources only and thus lack sophisticated resource management services. Coordinated use of resources at multiple sites is not supported at present in Grid systems. DGET provides a powerful resource management facility for a wide variety of resources. Security is not a major concern in P2P systems therefore most of the P2P systems lack a sophisticated security component. Most of the P2P systems focus on maintaining anonymity[12]. DGET employs a very sophisticated security mechanism based on the extended Java security model. P2P systems lack the support of migration capabilities but DGET is designed to efficiently migrate the Entities triggered by a change in the needs of the Entities, variation in the system conditions or as a result of load balancing or task scheduling.

*DGET and Hybrid Systems:* Some system designers have tried integrating both P2P and Grid approaches to come up with a system which enjoys the benefits of both Grid and P2P systems[4]. Such systems are known as hybrid P2P Grid systems. OurGrid is one such P2P Grid middleware. OurGrid[2] bears many similarities with DGET but has some differences as well. OurGrid lacks sophisticated resource discovery solutions present in DGET. Another difference between DGET and OurGrid is migration support. DGET supports strong transparent migration of applications but OurGrid does not.

## III. DGET Overview

As described in the related work section, there are a few systems that have combined the concepts from both Grid and P2P systems. Such hybrid systems are called P2P Grids. DGET adopts the same approach and exploits the advantages of both systems and provides an integrated environment for manipulating and analyzing very large data sets.

### A. DGET Objectives

We have set the following high-level objectives for the DGET middleware.

*Uniform Management Interface:* Resources in the DGET system are represented through a standard and uniform interface. This approach helps in masking the intra-resource heterogeneity. Users don't have to master the entire heterogeneous interface. New resources can be seamlessly added to the system.

*Simplicity & Ease of Use:* Grid users are typically non-technical, therefore, it is imperative that Grid middleware should be simple and easy to use. DGET should tackle the low-level complexities and make it simple for the Grid users to use and manage.

*Fault Tolerance:* In a large scale Grid system, faults are not an exception but a norm. DGET should be able to manage and survive system failures transparently without degrading the application performance.

*Scalable Architecture:* The DGET architecture should be scalable to accommodate thousands or even millions of users, resources and data sets. The DGET topology must be decentralized and dynamic as centralized architectures result in poor system scalability.

### B. DGET Concepts

*Entity:* An Entity is a network enabled discrete unit of abstraction that provides some functionality to its users. An Entity can take many forms e.g. a remote activity, a remote object, a server that processes user requests etc. The Concept of an Entity is akin to a process. An Entity is a mobile element that can move around on different Nuclei. An Entity is composed of two parts, a system provided Shell and user provided Ghost. Definitions of these are given below.

*Shell:* The Shell is the system provided control part of the Entity. The Shell exposes a management interface through which Entities can be manipulated. The Shell is attached to the programmer provided Ghost when an Entity is instantiated.

*Ghost:* The Ghost represents the programmer provided part of an Entity, it implements the actual logic or functionality of the Entity.

*Nucleus:* The Nucleus is the kernel of the system. It Provides basic services like lifecycle management, communication, security etc. to Entities.

*Connector:* Transport protocol agnostic communication medium provided to Entities for communicating with each other.

## IV. DGET Architecture

In this section we will give an overview of the architectural, components describing how they are structured and organised. Detailed descriptions of these components are given in their respective sections. Figure 1 shows a diagrammatic overview of the system. The DGET system is composed of three logical layers. The Following is a brief description of each layer and the components residing in that layer.

*Core Layer:* This layer provides basic services to the Entities executing in the Nucleus. These basic services include communication facilities, lifecycle management and security.

*Facilitation Layer:* This is the second layer in the system. It facilitates execution of the Entities by providing them certain services. The components residing at this layer are also modeled as Entities. Entities residing at this layer are called System Entities. System Entities use the services provided by the core layer. Certain components from the Core layer are modeled as System Entities as well. Therefore, in the diagram, the Security and EntityManager components span both Core and Facilitation layers. The following Entities are located at this layer:

- **Entity Manager Entity:** This Entity provides lifecycle management services, such as instantiating new Entities or manipulate existing Entities.

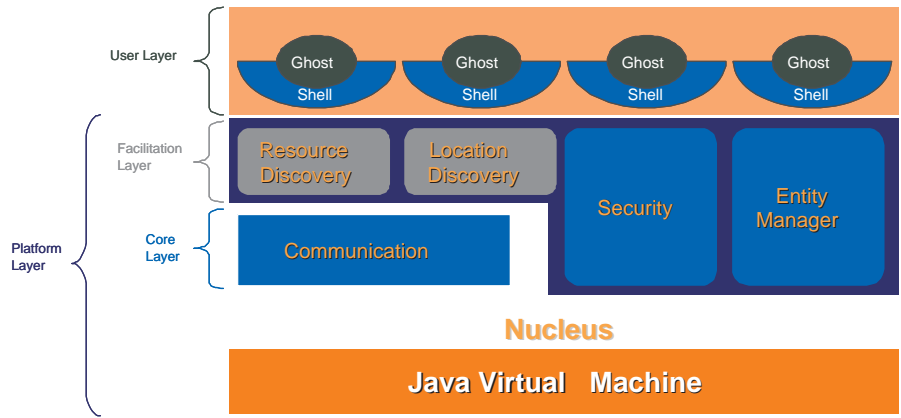

Fig. 1. DGET Architectural Compenents

- **Security Entity:** This Entity provides security related services to other Entities.
- **Resource Discovery Entity:** This Entity provides the service of discovering other Entities.
- **Location Discovery Entity:** This Entity works in conjunction with the Resource Discovery Entity and provides the service of discovering the location of other Entities.

*User Layer:* This is the top most layer in the system. Entities developed by the users and deployed into the system reside at this layer. Entities located at this layer provide user implemented functionality to the users.

## V. ENTITY MODEL

### A. Entity Make-up

As described above, an Entity is composed of two parts: a Shell and a Ghost. The Shell is the system provided part that is attached to the programmer supplied Ghost when an Entity is instantiated. The Shell supports operations to manipulate Entities during their execution. The list of methods exposed by the Shell is given below. Since a Shell is attached to every Entity executing in the Nucleus, all the Entities can be manipulated using this uniform management interface.

```
public class Shell {
    public void Shell(){...}
    protected void start(){...}
    public void stop(){...}
    public void suspend(){...}
    public void resume(){...}
    public void export(){...}
    protected void import(){...}
}
```

Besides providing a uniform management interface for Entities, the Shell serves other purposes as well. The Shell intercepts all the method invocations to enforce access control on behalf of the Ghost. This relieves programmers from implementing access control functionality in the Ghost logic. Access control policies are specified in the deployment descriptor and thus can be changed without changing the code of the Ghost.

The Ghost, the second part of the Entity that actually implements the Entity logic is provided by the programmer. The programmer writes the Ghost by extending the Ghost superclass provided by DGET. The Ghost class supports the following methods to access the system services provided by the Nucleus.

```
public class Ghost {
  protected final void setContext(){}
  public final EntityContext getContext(){}
  public final Shell getShell(){}
}
```

### B. Entity Types

Entities can be divided into the following categories based on their communication characteristics.

*Operation-Driven Entities:* These type of Entities exposes operations that are accessed by invoking methods on them. In other words, communication with them is typed. Method invocation is done transparently by the Shell attached to the Ghost. These type of Entities provide a programming framework, which users can use to develop Grid applications and deploy them on the Grid.

*Data-Driven Entities:* These type of Entities interact by exchanging messages. These Entities use un-typed communication. The purpose of introducing this category is to allow the development of custom protocols and thus new programming frameworks for DGET.

*Hybrid Entities:* Entities operating with both communication models can also exist in the system. An example of such an Entity could be an application that receives raw information from some source, processes it and allow its client to access typed information by invoking methods on it.

## VI. DGET IMPLEMENTATION

### A. Entity Management

*Entity Creation & Isolation:* The EntityManager component is responsible for initiating the instantiation of an Entity in the Nucleus. As previously described, the

`EntityManager` functionality is exposed as a System Entity in the Nucleus. The `EntityManager` Entity (EME) publishes its existence along with the characteristics of the host so other Entities can locate EMEs according to their requirements. In order to access the local EME running in the same Nucleus, Entities can use `EntityContext`. EME instantiates a Shell and passes it the system parameters required to load a Ghost. These system parameters include a `GhostLoader` reference, `ThreadGroup` reference and information about the Ghost to be instantiated. The Shell uses these parameters and instantiates the Ghost. After successful instantiation of the Ghost, the Shell calls the `setEntityContext()` method on the Ghost class passing in the `EntityContext` object. The Shell also passes an instance of itself to the Ghost. The Ghost can use this instance to invoke lifecycle management operations on itself.

The `EntityContext` class is the medium Ghosts can use to access system services supported by the Nucleus.

```
public class EntityContext {
  public Connector getEMEntity();
  public Connector getRDEntity(){};
  public Connector getLDEntity{};
  public Shell getShell();
  public NucleusInfo getNucleusInfo();
  public Resource[] getResourceLimits();
  public Resource[] getResourceConsumption();
}
```

The first three methods return connectors to the Entity Manager, Resource Discovery and Location Discovery System Entities running in the Nucleus. A Ghost can access an instance of its Shell through `EntityContext` as well. In addition to providing access to System Entities, the `EntityContext` class provides access to other information as well. This includes information about the Nucleus, resource limits for the Ghost and resource consumption of the Ghost.

Entity isolation in the Nucleus is provided by a custom classloader called `GhostLoader`. A separate GhostLoader is used to load all the classes belonging to an Entity thus providing a separate namespace for the Entity classes. The GhostLoader associates a security context with the Entities classes. This security context is used during its execution to take the access control decisions. Entities are not allowed to pass references to each other while communicating. Communication between Entities is done through connectors. The `GhostLoader` has other functions in DGET beside providing separate namespaces for Entity classes. These include instrumenting Entity bytecode to support soft termination, transparent migration and resource control etc. Figure 2 shows screenshots of the EntityManager GUI. The figure on the left is the Entity viewer tab. It displays a list of Entities running in the Nucleus. Entities can be stopped and migrated through this tab. The figure on the right displays the Entity launcher tab. New Entities can be launched through this interface.

*Soft Termination:* Entity termination means killing all the threads an Entity has created during its execution. Sun has declared the thread termination methods as potentially unsafe[13] and deprecated them. Another approach should be adopted to terminate all the threads belonging to an Entity. DGET uses the following approach for soft termination of Entity threads:

An `Execution` class is introduced. This `Execution` class has a flag indicating the execution state of the Entity. During the execution, all Entity threads call the `check()` method periodically. If the execution flag is RUNNING, the `check()` method returns silently but if execution state is TERMINATED, the `check()` method throws `EntityTerminatedException`. During the loading process, Entity classes are instrumented with these execution checks. All the methods are also instrumented with `try/catch` blocks. The `catch` catches the `EntityTerminatedException` exception and re-throws this exception to propagate it down the thread stack. Entity classes are not allowed to catch this exception. During the classloading and instrumentation process, Entity class files are scanned to find exception handlers for the `EntityTerminatedException`. This scanning ensures that malicious programmer don't catch this error in order to prevent the Entity termination.

*B. Distributed Authentication*

For the DGET middleware we decided to implement an Identity Based Authentication system that will enable the middleware to scale and provide fine grained administration of the authentication and authorisation scheme. We decided to implement such a system because the standard PKI infrastructure used in Grid systems has many drawbacks related to the complexity involved in managing certificates, revocation lists and various cross-certification problems. This complexity leads to a lot of administrative overhead and expensive infrastructure, which has prevented PKI technology from enabling truly ubiquitous secure messaging and authentication. This problem is complicated in Grid systems such as Globus because we have to add support for delegation of capabilities as shown in [14]. The two majors features provided through Grid Security Infrastructure (GSI) [15] is Single Sign On (SSO) and unattended user authentication (jobs can have a very long execution time and spawn a multitude of sub processes, requiring human intervention is not feasible). To enable both features GSI relies heavily on the X509 certificate and certificate chain, revocation list. But as the number of users, resources and requirements evolve the complexity of such a solution expands rapidly while its scalability diminishes. More information means more complex policies. More policies mean larger Certificate Revocation Lists. Larger CRLs mean more management and more overhead.

We decided to design an Identity Based Cryptographic (IBC) solution to handle authentication in DGET. We were inspired by various solutions that appeared lately such as [14], [16], [17]. It provides an easy way to manage keys and the benefits of the ID-based approach include:

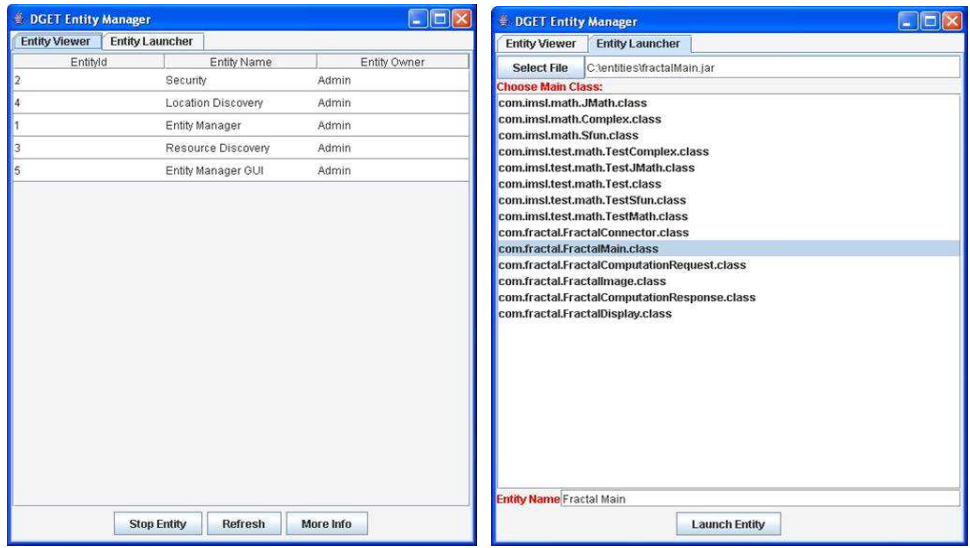

Fig. 2. DGET Entity Manager GUI

- Automatic revocation via expiry of time-limited identifiers.
- Reduced round trips if the user can predict delegation requests.
- Reduced bandwidth.
- Similar computational costs.
- Trivial computation of proxy key pairs (RSA key pair generation replaced by elliptic curve multiplication).
- Replication of existing Grid security capabilities.

There are some drawbacks as well

- Timestamp automatic revocation requires the setup of a complex reissue systems for long delegation chains.
- The key escrow system can be problematic for cross administration/company Grids.

So far we have implemented the IBC starting from the Identity Based Encryption (IBE) java library developed by the University of Maynooth [18] where we added the SOK-IBS ( Sakai-Ogishi-Kasahara IdentityBased Signature) capability. We are now in the phase of integrating the IBC within the core of DGET and the network stack.

## C. Migration Support

One of the distinguishing features of DGET is transparent strong migration support. This section describes implementation details of migration support in DGET.

*1) Implementation Techniques:* A brief overview of the implementation techniques used in the migration solution is given here. A deciding factor in choosing these methodologies were the requirements of portability of the solution, minimal space and time performance overhead. Bytecode Instrumentation is the first technique to support transparent migration. Entity classes are instrumented and bytecode blocks are injected. These injected code blocks perform different functions like program counter restoration and execution checkpoints (described shortly). This bytecode instrumentation is performed at class load time by a custom classloader. Bytecode instrumentation is performed by the classloader using the Byte Code Engineering Library(BCEL)[19]. The second technique used for capturing and restoring execution state is the Java Platform Debugger Architecture (JPDA). JPDA is part of the JVM specification and thus it is implemented by every standard JVM implementation. JPDA provides access to runtime information of the JVM including the thread stacks. JPDA is implemented purely in Java so our migration solution doesn't lose portability

*2) Migration Enabling Features:* I will now explain the features that enable transparent strong migration in DGET. In order to perform migration at an arbitrary point, values on the operand stack must be saved and restored during the Entity restoration process. Unfortunately, JPDA doesn't expose any methods to access the values currently present on the operand stack. Initiating migration in the middle of a source code level instruction might result in loss of data from the operand stack. One solution could be to insert checkpoints in the code at locations where execution is not in the middle of a source code level instruction. Migration requests should be delayed until the execution reaches any such checkpoint. Execution checkpoints are inserted as the first instruction in every method and in all the loops in every method.

Another DGET feature to support multi-threaded migration is Mobile Monitors. Java provides multi-threading support in the form of `synchronized` methods and code blocks. A monitor is associated with each java object by the JVM and before entering a `synchronized` method or code block, a thread has to acquire the monitor associated with the object. Monitors associated with java objects are maintained and hidden inside the JVM. These monitors are not serializable and thus are not transported with the serialized objects. Mobile Monitors preserve the lock state upon migration. During the class loading process, class files are instrumented to replace Synchronized methods and code blocks. A Mobile Monitor is associated with a class that requires synchronized access.

*3) Migration Process:*

*Entity Suspension:* The migration process is initiated when the `export()` method is invoked on the Shell. Before the execution state can be captured, all the running threads of the Entity must be suspended. Sun Microsystems has deprecated[13] `Thread.suspend()` and programmers are instructed to use alternative measures. Execution checkpoints discussed in the previous section are used to halt the execution of the Entity so its execution state can be captured. The `export` method calls the `suspend()` method on the associated `Execution` class. As a result, execution of all the threads is blocked on the next execution checkpoint. At this point, all the Entity threads would be in one of the following states:

1) Execution wait set: Thread reached execution checkpoint and blocked as a result of `check()` method call.
2) Monitor entry set: Thread tried to acquire a lock and was blocked as the lock was held by another thread.
3) Monitor wait set: Thread called the wait() method inside a synchronized method or block and is blocked waiting for the `notify()` or `notifyAll()` method to be invoked by another thread.
4) Thread called the `join()` method on another thread and is waiting for the thread to finish execution.
5) Thread called `sleep()` method and is sleeping.

*State Capture:* After the execution of all the Entity threads has been halted, execution state capture can be started. JPDA, discussed previously, is used to capture the execution state of all the Entity threads. The `StackFrame` class from JPDA represents a method call on the thread stack. The `StackFrame` class gives access to the values of local variables and the program counter. Each local variable is represented as the `LocalVariable` class. Calling the `visibleVariable()` method on the `StackFrame` class returns a list of all the variables accessible till the point of execution in the method code. The `location()` method returns the `Location` class that represents the location in the stack frame. The `codeIndex()` method can be used to extract the code index relative to the start instruction of the method. Using these classes and methods, the execution state of all the methods on the thread stack can be accessed and saved. The execution state of all the Entity threads along with the Mobile Monitors and Execution class is saved in a serializable format and transported to the destination for reincarnation of the Entity.

*State Restoration:* On the destination Nucleus, the Entity state is restored by calling the `import()` method of the Shell. The saved image of the Entity's execution context is passed as a parameter to the `import()` method. The Entity's threads must be launched in special order to avoid any race conditions. Threads blocked as a result of the `wait()` method call must be launched first, followed by the threads blocked on the `Execution` class.

To reestablish the execution state of a thread, its method stack must be rebuilt. To do this, all the methods are called in the order they were on the stack before execution was suspended and migration was initiated. Event handlers can be set that are called when method entry/exit event occurs. When a method entry event occurs, such event handlers restore the values of local variables of the method from the saved execution image. After restoring the local variables execution should continue from the code position which is the method invocation for the next method on the stack. Doing so will ensure the instructions already executed are skipped and restoration of the next method on the stack frame begins and proceeds in the same manner. After restoring all the threads to the state they were before the migration was initiated, the `resume()` method of the `Execution` class is called. This method sets the execution status flag to RUNNING and notifies all the threads blocked on this class. Execution will proceed normally afterwards.

As mentioned in the previous paragraph, after restoring local variables, execution should continue from the code position which is the method invocation of the next method on the stack. No mechanism is available in JPDA to set the value of the program counter register to this code position. This problem is solved by maintaining an Artificial Program Counter (APC) which represents an index of method invocations in the method. This APC is incremented after every method invocation instruction. This APC is used in conjunction with a `tableswitch` bytecode instruction which branches the execution according the value of the APC. This `tableswitch` and APC increment instructions are added during the instrumentation procedure. The `tableswitch` instruction is added at the beginning of each method and defaults to the original starting code position of the method code.

## VII. CONCLUSION

We have described the architecture of the DGET middleware for P2P Grid systems. DGET simplifies the deployment, management and usage of Grid systems. DGET provides a dynamic and scalable solution that relies on a truly decentralized and self-organizing topology. DGET enables resource sharing with the least management overhead and eases the task of Grid programming. In the future, we plan to incorporate more sophisticated features like fine-grained resource control thus making it feasible to provide Quality of Service (QoS) and support and enforce Service Level Agreements (SLA).